\documentclass[10pt,leqno]{amsart}
\usepackage[foot]{amsaddr}
\usepackage{graphicx}
\usepackage{indentfirst,csquotes}

\usepackage{amssymb,amsthm,amsmath}
\usepackage{xcolor,paralist,hyperref,titlesec,fancyhdr,etoolbox}
\usepackage{subcaption}
\usepackage{adjustbox}

\titleformat{\section}[display]{\normalfont\huge\bfseries\centering}{\centering\chaptertitlename\thechapter}{10pt}{\Large}
\titlespacing*{\section}{0pt}{0ex}{0ex}

\hypersetup{ colorlinks=true, linkcolor=black, filecolor=black, urlcolor=black }

\usepackage{lipsum}

\begin{document}
\title{Ternary superconducting hydrides in the La--Mg--H system} 
\author[]{Grigoriy Shutov$^1$}
\author[]{Dmitrii Semenok$^2$}
\author[]{Ivan Kruglov$^{3, 4}$}
\author[]{Artem Oganov$^1$}
\date{\today}
\address{$^1$Skolkovo Institute of Science and Technology, Moscow, Russia}
\address{$^2$Center for High Pressure Science and Technology Advanced Research (HPSTAR), Beijing, China}
\address{$^3$Moscow Institute of Physics and Technology, Moscow, Russia}
\address{$^4$Dukhov Research Institute of Automatics (VNIIA), Moscow, Russia}
\email{Grigoriy.Shutov@skoltech.ru}
\maketitle


\begin{abstract}
Ternary or more complex hydrogen-rich hydrides are the main hope of reaching room-temperature superconductivity at high pressures. Their chemical space is vast and its exploration is challenging. Here we report the investigation of the La–Mg–H ternary system using the evolutionary algorithm USPEX at pressures on the range 150-300 GPa. Several ternary superconducting hydrides were found, including thermodynamically stable $P6/mmm$-LaMg$_{3}$H$_{28}$ with $T_{\mathrm{C}}=164$~K at 200~GPa, $P/2m$-LaMgH$_8$, $C2/m$-La$_2$MgH$_{12}$ and $P2/m$-La$_3$MgH$_{16}$. In addition, novel binary hydrides were predicted to be stable at various pressures, such as $Cm$-Mg$_6$H$_{11}$, $P1$-MgH$_{26}$, $Fmm2$-MgH$_{30}$, $P1$-MgH$_{38}$ and $R\overline{3}m$-LaH$_{13}$. We also report several novel low-enthalpy metastable phases, both ternary and binary ones. Finally, we demonstrate important methods of exploring very large chemical spaces and show how they can improve crystal structure prediction. 
\end{abstract} 

\bigskip

\section*{Introduction}
\label{sec:introduction}

\par The highest-temperature superconductors know today are polyhydrides - hydrides anomalously rich in hydrogen (above what can be expected based on atomic valences). Their studies were anticipated with the hypothesis proposed by Ashcroft \cite{Ashcroft1968} that hydrogen-rich materials can become superconductors at high pressures. Some hydride systems demonstrate high superconducting critical temperature, such as $Im\overline{3}m$-$\mathrm{H_3S}$ with $T_{\mathrm{C}}=203$~K \cite{Drozdov2015}, $Fm\overline{3}m$-$\mathrm{LaH_{10}}$ with $T_{\mathrm{C}}=250$~K \cite{Drozdov2019, Somayazulu2019}, $P6_3/mmc$-ThH$_9$ and $Fm\overline{3}m$-ThH$_{10}$ with $T_{\mathrm{C}}=146$~K and $T_{\mathrm{C}}=159-161$~K \cite{th-h}, respectively, and $Im\overline{3}m$-YH$_6$ with $T_{\mathrm{C}}=224-226$~K \cite{yh6}.

\par Semenok et al. \cite{Semenok2020} showed that high-$T_{\mathrm{C}}$ superconducting hydrides are formed by elements of II-III groups, such as calcium, strontium, barium, scandium, yttrium, lanthanum and lutetium. Elements grouped near these (to a lesser extent near sulfur) in the periodic table often form binary hydrides with the predicted $T_{\mathrm{C}}>100$~K. 

\par The search for polyhydrides continues in ternary systems. Ternary hydrides $Fm\overline{3}m$-(La,Y)H$_{10}$ have been synthesized, and $T_{\mathrm{C}}=253$~K has been reported~\cite{layh}. Lithium phosphorus hydride $Pm\overline{3}$-LiPH${_6}$ has been predicted to be a superconductor with $T_{\mathrm{C}} = 150$–$167$~K~\cite{Shao2019}. 

\par As $T_{\mathrm{C}} = 250$~K of LaH$_{10}$ has been proven experimentally~\cite{Drozdov2019, Somayazulu2019}, we can presume that ternary hydrides of lanthanum and some other element may also be high-temperature superconductors. In this work, we study phases formed by lanthanum, magnesium and hydrogen and their superconducting properties.

\par Lanthanum hydrides have been theoretically predicted to form several phases at high pressures: $P6/mmm$-$\mathrm{LaH_{2}}$, $Cmmm$-$\mathrm{La_{3}H_{10}}$, $I4/mmm$-$\mathrm{LaH_{4}}$, $Fm\overline{3}m$-$\mathrm{LaH_{10}}$, and $P6/mmm$-$\mathrm{LaH_{16}}$~\cite{Kruglov}. Moreover, experimental synthesis revealed 7 phases of binary lanthanum hydrides \cite{laniel2022high}. According to predictions, $\mathrm{LaH_{10}}$ is a superconductor with $T_{\mathrm{C}} = 274–286$~K \cite{liu2017potential}. Magnesium hydrides also form four phases at 200~GPa: $P6_3/mmc$-$\mathrm{MgH_{2}}$, $Cmcm$-$\mathrm{MgH_{4}}$, $R\overline{3}$-$\mathrm{MgH_{12}}$, and $P1$-$\mathrm{MgH_{16}}$. However, they have relatively low reported $T_{\mathrm{C}}$: 29-37~K for $\mathrm{MgH_{4}}$ and 20~K for $\mathrm{MgH_{2}}$ and 47-60~K for $\mathrm{MgH_{12}}$. For $\mathrm{MgH_{16}}$, $T_{\mathrm{C}}$ has not been calculated~\cite{Lonie2013}. Another study reports $Im\overline{3}m$-MgH$_6$ with $T_{\mathrm{C}}=260$ K at pressure above 300 GPa \cite{Feng2015}.

\par In this work, we use the evolutionary algorithm USPEX~\cite{Oganov2006, Oganov2011, Lyakhov2013} to study the chemical space of the $\mathrm{La}$–$\mathrm{Mg}$–$\mathrm{H}$ system. For more detailed search of La-Mg-H ternary hydrides, we additionally perform evolutionary searches on the pseudobinary sections formed by Mg-H and La-H binary hydrides.

\section*{Methods}
\label{sec:methods}

\par The evolutionary algorithm USPEX~\cite{Oganov2006, Oganov2011, Lyakhov2013} was used to predict thermodynamically stable phases. To investigate the $\mathrm{La}$–$\mathrm{Mg}$–$\mathrm{H}$ system, we performed both fixed- and variable-composition searches at 200~GPa.

\par One of the possible straightforward ways to search for new stable hydrides in La-Mg-H system is to perform USPEX calculations in the ternary system. However, such search requires large computational resources, because the chemical space of ternary systems is huge. To tackle this problem, we performed variable-composition searches on special pseudobinary sections of the convex hull with LaH$_{\mathrm{x}}$--MgH$_{\mathrm{y}}$ ($x=2, 4, 10, 16$, $y=2, 4, 12, 16$) composition blocks. In addition, we performed variable-composition searches with such composition blocks as LaH$_{\mathrm{x}}$--Mg and La--MgH$_{\mathrm{y}}$ ($x=2, 4, 10, 16$, $y=2, 4, 12, 16$ as well).  The parameters of each search are presented in Supplementary Materials, Table S1.

\par After studying the pseudobinary sections, the ternary variable-composition search at 200 GPa pressure was performed, using previously found structures as seeds. The number of generations was 100. After this search, several ternary hydrides remained on the convex hull. Moreover, novel binary hydrides were discovered. Using stable and metastable structures at pressure of 200 GPa as seeds, ternary convex hulls were also calculated at 150, 250 and 300 GPa. Additionally, we recalculated these ternary convex hulls on the temperature range from 0 K to 2000 K, using free energies computed by Phonopy. Metastable structures with the energy above hull $E_{\mathrm{Hull}} \leq 10$~ meV/atom are also presented on each convex hull in this work. 

\par Structure relaxations and energy estimation were performed using the VASP code~\cite{Kresse1996, Kresse1995, Kresse1994} within density functional theory (DFT)~\cite{Hohenberg1964, Kohn1965}, implementing the Perdew–Burke–Ernzerhof (PBE) exchange–correlation functional~\cite{Perdew1996} and the projector-augmented wave (PAW) method~\cite{Bloch1994, Kresse1999}. The kinetic energy cutoff was set at 600~eV. $\Gamma$-centered $k$-point meshes with a resolution of $2\pi \times 0.05$~Å$^{-1}$ were used for sampling the Brillouin zone.

\par The phonon band structure and density of states were computed using Phonopy~\cite{phonopy} package implementing the finite displacement method. $2\times2\times2$ supercells were generated. The energy cutoff and $k$-spacing parameters for the VASP calculations were set at 500~eV and $2\pi\times0.1$~Å$^{-1}$, respectively. Sumo package~\cite{sumo} was used to visualize the phonon density of states and band structure. The k-points for phonon band structures were chosen using Hinuma's recommendation~\cite{Hinuma2017}. The Phonopy package was also used to calculate zero-point energy (ZPE) corrections and thermal properties, such as the entropy and free energy. In addition, structures' symmetries were also investigated by the Phonopy package. Some of the predicted structures can be symmetrized to various space groups depending on the tolerance parameter. We chose the maximum symmetry space groups within whose structures exhibited dynamical stability. 

\par To calculate phonon frequencies and electron–phonon coupling (EPC) coefficients, we used Quantum Espresso (QE) package~\cite{Giannozzi_2009} utilizing density functional perturbation theory (DFPT)~\cite{Baroni2001}, plane-wave pseudopotential method, and the PZ-HGH~\cite{pz, pz-hgh} pseudopotentials. The $q$-meshes for each structure were $3\times3\times3$, except MgH$_{26}$, MgH$_{30}$ and MgH$_{38}$, for which it was set to $2\times2\times2$. The Allen–Dynes~\cite{Allen1973} formula was used to calculate $T_{\mathrm{C}}$: 

\begin{equation}
    T_{\mathrm{C}}=\omega_{\log } \frac{f_{1} f_{2}}{1.2} \exp \left[\frac{-1.04(1+\lambda)}{\lambda-\mu^{*}-0.62 \lambda \mu^{*}}\right].
    \label{mad}
\end{equation}
\par The McMillan formula has the term $f_1f_2=1$, whereas in the Allen–Dynes formula, it is expressed as:
\begin{align}
    f_{1} f_{2}&=\sqrt[3]{1+\left[\frac{\lambda}{2.46\left(1+3.8 \mu^{*}\right)}\right]^{3 / 2}} \nonumber \\  &\cdot\left[1-\frac{\lambda^{2}\left(1-\omega_{2} / \omega_{\log }\right)}{\lambda^{2}+3.312\left(1+6.3 \mu^{*}\right)^{2}}\right],
\end{align}
where $\mu^*$ is the Coulomb pseudopotential, with the values in the commonly accepted range from 0.10 to 0.15; $\lambda$, $\omega_2$, and $\omega_{\log}$ are the EPC constant, mean square frequency, and logarithmic average frequency, respectively, defined as:
\begin{equation}
    \lambda=\int_{0}^{\omega_{\max }} \frac{2 \alpha^{2} F(\omega)}{\omega} \mathrm{d} \omega,
\end{equation}
\begin{equation}
    \omega_{\log }=\exp \left[\frac{2}{\lambda} \int_{0}^{\omega_{\max }} \frac{\mathrm{d} \omega}{\omega} \alpha^{2} F(\omega) \log (\omega)\right],
\end{equation}
\begin{equation}
    \omega_{2}=\sqrt{\frac{1}{\lambda} \int_{0}^{\omega_{\max }}\left[\frac{2 \alpha^{2} F(\omega)}{\omega}\right] \omega^{2} \mathrm{~d} \omega}.
\end{equation}
where $\alpha^{2} F(\omega)$ is an Eliashberg function. 

\par Moreover, full solution of Eliashberg equations \cite{Eliashberg1960} was computed using  Allen's algorithm \cite{Allen1975}. This algorithm was implemented in our code published on GitHub \cite{GitGreg228}. 

\par To estimate the thermodynamic properties such as critical magnetic field $H_{\mathrm{C}}$, superconducting gap $\Delta(0)$ and specific heat jump $\Delta C(T_{c})$, we used the semiempirical formulas~\cite{carbotte1990}:
\begin{equation}
    \frac{\gamma T_{\mathrm{C}}^{2}}{H_{\mathrm{C}}^{2}(0)}=0.168\left[1-12.2\left(\frac{T_{\mathrm{C}}}{\omega_{\log }}\right)^{2} \ln \left(\frac{\omega_{\log }}{3 T_{\mathrm{C}}}\right)\right],
    \label{eq1}
\end{equation}
\begin{equation}
    \frac{2 \Delta(0)}{k_{\mathrm{B}} T_{\mathrm{C}}}=3.53\left[1+12.5\left(\frac{T_{\mathrm{C}}}{\omega_{\log }}\right)^{2} \ln \left(\frac{\omega_{\log }}{2 T_{\mathrm{C}}}\right)\right],
    \label{eq2}
\end{equation}
\begin{equation}
    \frac{\Delta C\left(T_{c}\right)}{\gamma T_{c}}=1.43\left[1+53\left[\frac{T_{c}}{\omega_{\log }}\right]^{2} \ln \left[\frac{\omega_{\log }}{3 T_{c}}\right]\right],
    \label{eq3}
\end{equation}
where $k_\mathrm{B}$ is the Boltzmann constant and $\gamma=\frac{2}{3} \pi^{2} k_{\mathrm{B}}^{2} N(0)(1+\lambda)$ is the Sommerfeld constant. The values of $T_{\mathrm{C}}$ from the solution of Eliashberg equations were used in these formulas.
\par We note that lower and upper critical magnetic fields $H_{\mathrm{C}1}$ and $H_{\mathrm{C}2}$ are related to $H_{\mathrm{C}}$ by the following equation:
\begin{equation}
    H_{\mathrm{C}} = \frac{\sqrt{H_{\mathrm{C}1} H_{\mathrm{C}2}}}{\sqrt{\ln \kappa}}
\end{equation}
where $\kappa$ is Ginzburg-Landau parameter, and that $H_{\mathrm{C}2}$ is usually one or two orders of magnitude higher than $H_{\mathrm{C}}$. For example, $\kappa = 20$, $H_{\mathrm{C}1}=0.55$ T and $H_{\mathrm{C}2}=143.5$ T were reported for $Fm\overline{3}m$-LaH$_{10}$ \cite{minkov2022magnetic}, leading to $H_{\mathrm{C}}=5.13$ T. Formula (\ref{eq1}) returns $H_{\mathrm{C}} = 6.7$ T, which is close to the experimental value.

\newpage

\section*{Results}

\begin{figure}[h!!!]
    \centering
    \includegraphics[width=0.99\linewidth]{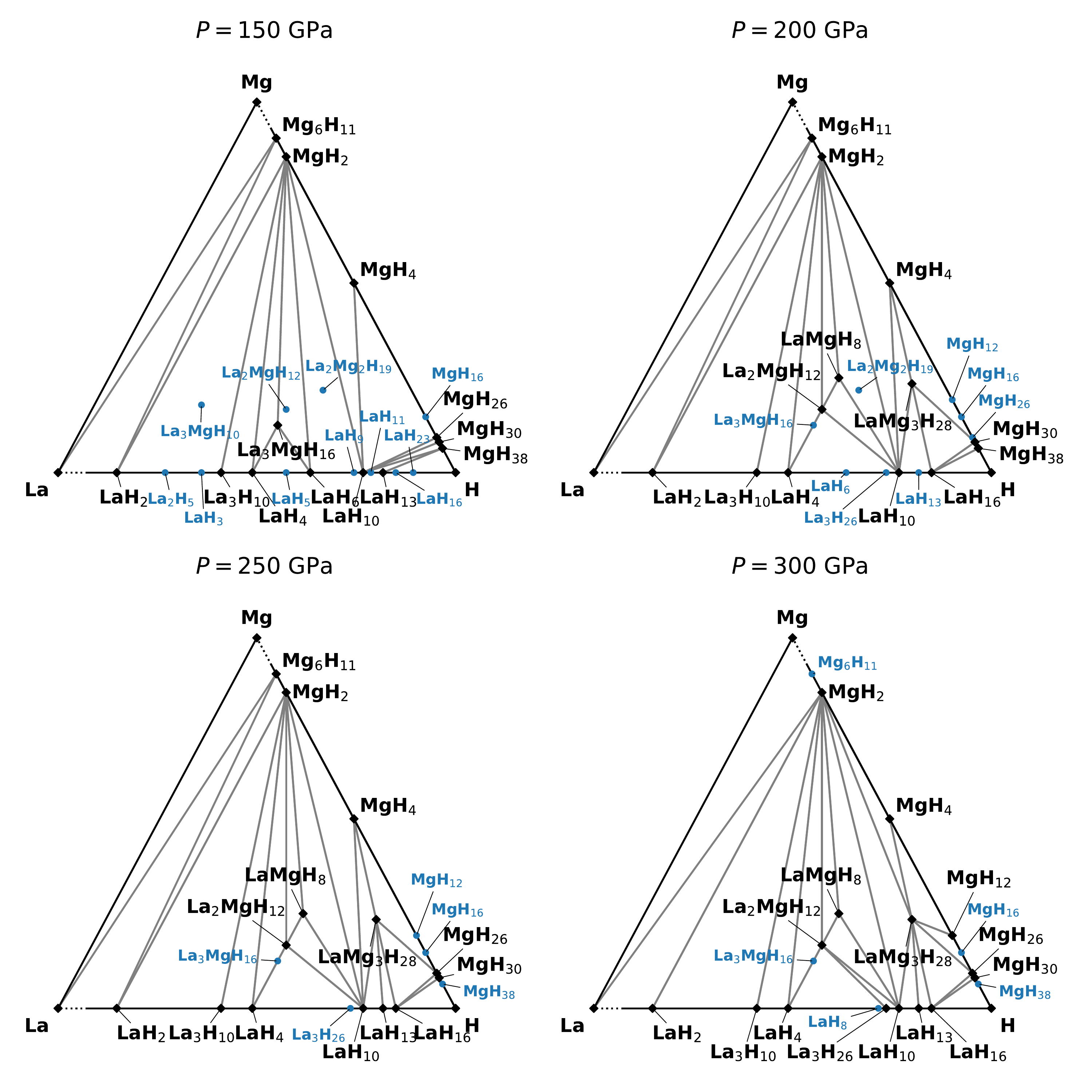}
    \caption{Convex hulls and phase diagrams of the La-Mg-H system, recalculated with zero-point energy corrections. The black diamonds are stable structures, the blue circles are metastable ones with the $E_{\mathrm{Hull}} \leq 10$~ meV/atom.}  
    \label{fig:chs}
\end{figure}

\par As a result of the variable-composition searches with LaH$_{\mathrm{x}}$--MgH$_{\mathrm{y}}$ ($\mathrm{x}=2, 4, 10, 16$, $\mathrm{y}=2, 4, 12, 16$) composition blocks at 200 GPa, 13 ternary phases were found to be stable with respect to their pseudoninary section (see Table S1). After variable-compostion searches on LaH$_{\mathrm{x}}$--Mg and La--MgH$_{\mathrm{y}}$ ($\mathrm{x}=2, 4, 10, 16$, $\mathrm{y}=2, 4, 12, 16$ as well) pseudobinary sections, 20 ternary phases were discovered, and some of them were metastable on the ternary convex hull. Other thermodynamically stable phases for each pseudobinary section are presented in Table~S1.

\par Ternary convex hulls and phase diagrams, recalculated with ZPE correction on $P=150$, 200, 250 and 300 GPa, are presented in Figure \ref{fig:chs}. They revealed $P/2m$-LaMgH$_8$, $C2/m$-La$_2$MgH$_{12}$ and $P6/mmm$-LaMg$_{3}$H$_{28}$ that are stable at $P=200$, 250 and 300 GPa and $P2/m$-La$_3$MgH$_{16}$ that is stable at $P=150$ GPa. 
\par Additionally, ternary variable-composition search at 200 GPa revealed novel binary lanthanum and magnesium hydrides: $R\overline{3}m$-LaH$_{13}$, $C2/m$-LaH$_{23}$, $Cm$-Mg$_{6}$H$_{11}$, $P1$-MgH$_{26}$, $Fmm2$-MgH$_{30}$ and $P1$-MgH$_{38}$.

\par Convex hulls and phase diagrams at non-zero temperatures are presented in Supplementary materials (see Figure S1-S4). Crystal structures are presented in Table S2.

\subsection*{Novel binary hydrides}

\begin{table*}[!!!htb]
\centering
\caption{Superconducting parameters of novel binary hydrides}
\adjustbox{max width=0.8\textwidth}{
\begin{tabular}{c|ccc|cc} \hline
Parameter & \begin{tabular}[c]{@{}c@{}}LaH$_{13}$\\ 200 GPa\end{tabular} & \begin{tabular}[c]{@{}c@{}}LaH$_{13}$\\ 250 GPa\end{tabular} & \multicolumn{1}{c|}{\begin{tabular}[c]{@{}c@{}}LaH$_{13}$\\ 300 GPa\end{tabular}} & \begin{tabular}[c]{@{}c@{}}LaH$_{23}$\\ 150 GPa\end{tabular} & \begin{tabular}[c]{@{}c@{}}LaH$_{23}$\\ 200 GPa\end{tabular} \\ \hline
$\lambda$ & 1.42 & 1.69 & 1.60 & 0.97 & 1.11 \\
$\omega_{\mathrm{log}}$, K & 906 & 812 & 1105 & 1195 & 1059 \\
$\omega_2$, K & 1454 & 1505 & 1669 & 1819 & 1844  \\
$T_{\mathrm{C}}$ (McM), K & 98 & 103 & 134 & 79 & 86 \\
$T_{\mathrm{C}}$ (A-D), K & 118 & 137 & 164 & 87 & 99 \\
$T_{\mathrm{C}}$ (E), K & 131 & 155 & 172 & 85 & 101 \\
$N_f$, $\frac{\mathrm{states}}{\mathrm{Ry}\cdot\texttt{\AA}^3}$ & 0.091 & 0.097 & 0.101 & 0.091 & 0.096 \\
$\gamma$, $\frac{\mathrm{mJ}}{\mathrm{cm}^3\cdot\mathrm{K}^2}$ & 0.126 & 0.150 & 0.152 & 0.103 & 0.116 \\
$\frac{\Delta C}{T_{\mathrm{C}}}$, $\frac{\mathrm{mJ}}{\mathrm{cm}^3\cdot\mathrm{K}^2}$ & 0.347 & 0.445 & 0.430 & 0.208 & 0.265 \\
$\Delta(0)$, meV & 26.4 & 34.0 & 35.6 & 14.5 & 18.2 \\
$\frac{2\Delta(0)}{k_{\mathrm{B}}T_{\mathrm{C}}}$ & 4.67 & 5.08 & 4.78 & 3.97 & 4.19 \\
$H_{\mathrm{C}}(0)$, T & 1.7 & 2.2 & 2.6 & 1.0 & 1.1 \\
\begin{tabular}[c]{@{}c@{}}Electron transfer,\\ e per H atom\end{tabular} & 0.231 & 0.231 & 0.231 & 0.130 & 0.130
\end{tabular}
}
\adjustbox{max width=0.65\textwidth}{
\begin{tabular}{c|c|c|c|c} \hline
Parameter & \begin{tabular}[c]{@{}c@{}}Mg$_6$H$_{11}$\\ 200 GPa\end{tabular} & \begin{tabular}[c]{@{}c@{}}MgH$_{26}$\\ 200 GPa\end{tabular} & \begin{tabular}[c]{@{}c@{}}MgH$_{30}$\\ 200 GPa\end{tabular} & \begin{tabular}[c]{@{}c@{}}MgH$_{38}$\\ 200 GPa\end{tabular} \\ \hline
$\lambda$ & 0.53 & 0.55 & 0.59 & 0.51 \\
$\omega_{\mathrm{log}}$, K & 898 & 1282 & 979 & 1002 \\
$\omega_2$, K & 1332 & 2273 & 2179 & 2167 \\
$T_{\mathrm{C}}$ (McM), K & 14 & 22 & 22 & 13 \\
$T_{\mathrm{C}}$ (A-D), K & 14 & 23 & 23 & 14 \\
$T_{\mathrm{C}}$ (E), K & 14 & 23 & 20 & 12 \\
$N_f$, $\frac{\mathrm{states}}{\mathrm{Ry}\cdot\texttt{\AA}^3}$ & 0.058 & 0.071 & 0.064 & 0.054 \\
$\gamma$, $\frac{\mathrm{mJ}}{\mathrm{cm}^3\cdot\mathrm{K}^2}$ & 0.051 & 0.063 & 0.059 & 0.047 \\
$\frac{\Delta C}{T_{\mathrm{C}}}$, $\frac{\mathrm{mJ}}{\mathrm{cm}^3\cdot\mathrm{K}^2}$ & 0.076 & 0.095 & 0.090 & 0.069 \\
$\Delta(0)$, meV & 2.2 & 3.5 & 3.2 & 1.8 \\
$\frac{2\Delta(0)}{k_{\mathrm{B}}T_{\mathrm{C}}}$ & 3.57 & 3.58 & 3.59 & 3.55 \\
$H_{\mathrm{C}}(0)$, T & 0.1 & 0.2 & 0.2 & 0.1 \\
\begin{tabular}[c]{@{}c@{}}Electron transfer,\\ e per H atom\end{tabular} & 1.091$^*$ & 0.077 & 0.067 & 0.053 
\end{tabular}
}
\caption*{$T_{\mathrm{C}}$ was calculated using McMillan (McM), Allen-Dynes (A-D) formulas, and numerical solution of Eliashberg equations (E) with Coulomb pseudopotential $\mu^*=0.1$.\\ $^*$one hydrogen atom can accept not more than one electron. Numbers greater than 1 indicate the presence of bonds between metal atoms.}
\label{tab:sclah}
\end{table*}

\begin{figure}[h!!!]
    \centering
    \begin{subfigure}[b]{0.35\linewidth}
    \includegraphics[width=0.99\linewidth]{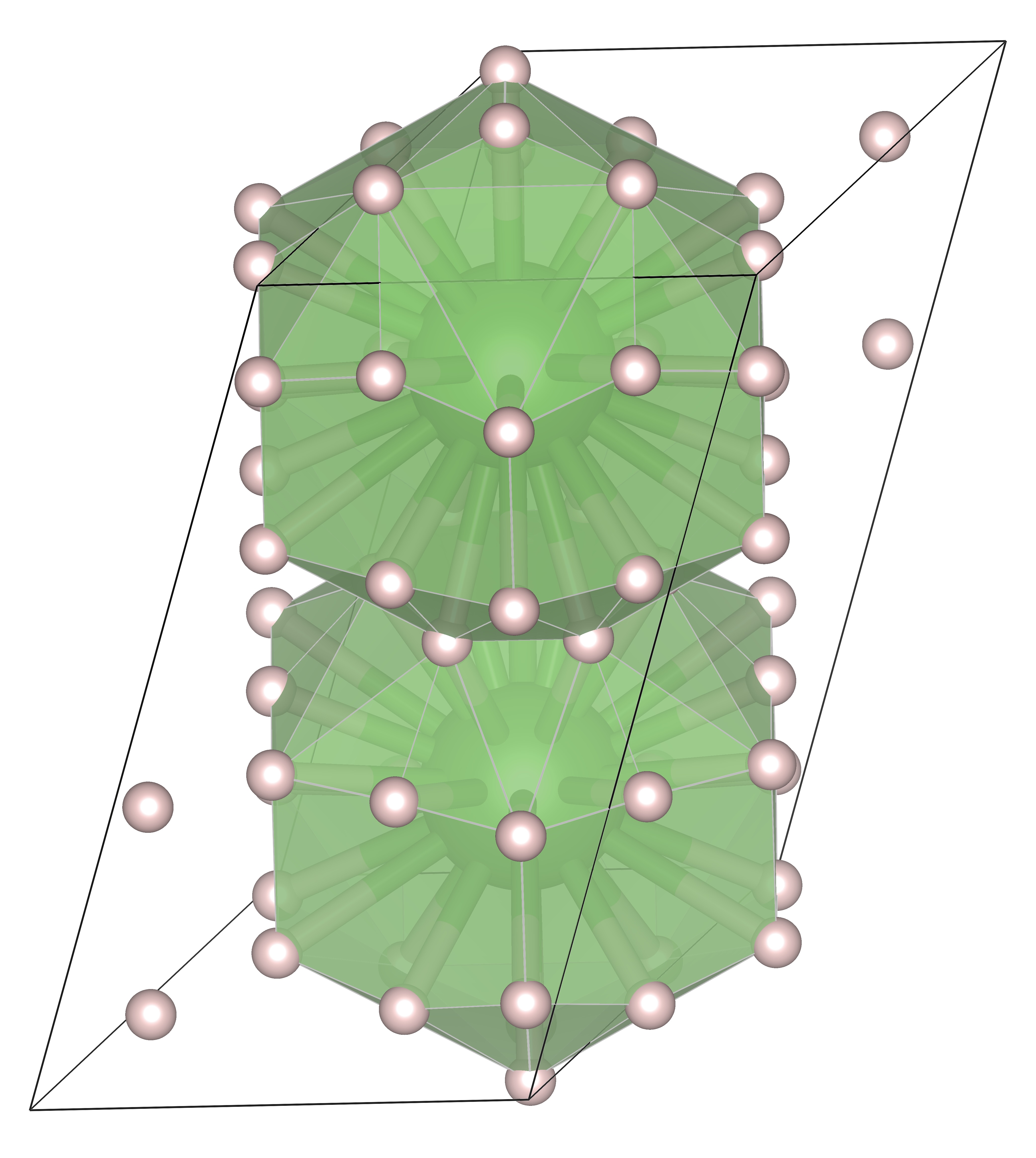}
    \caption{$R\overline{3}m$-LaH$_{13}$}
    \label{fig:lah13}
    \end{subfigure}
    \begin{subfigure}[b]{0.5\linewidth}
    \includegraphics[width=0.99\linewidth]{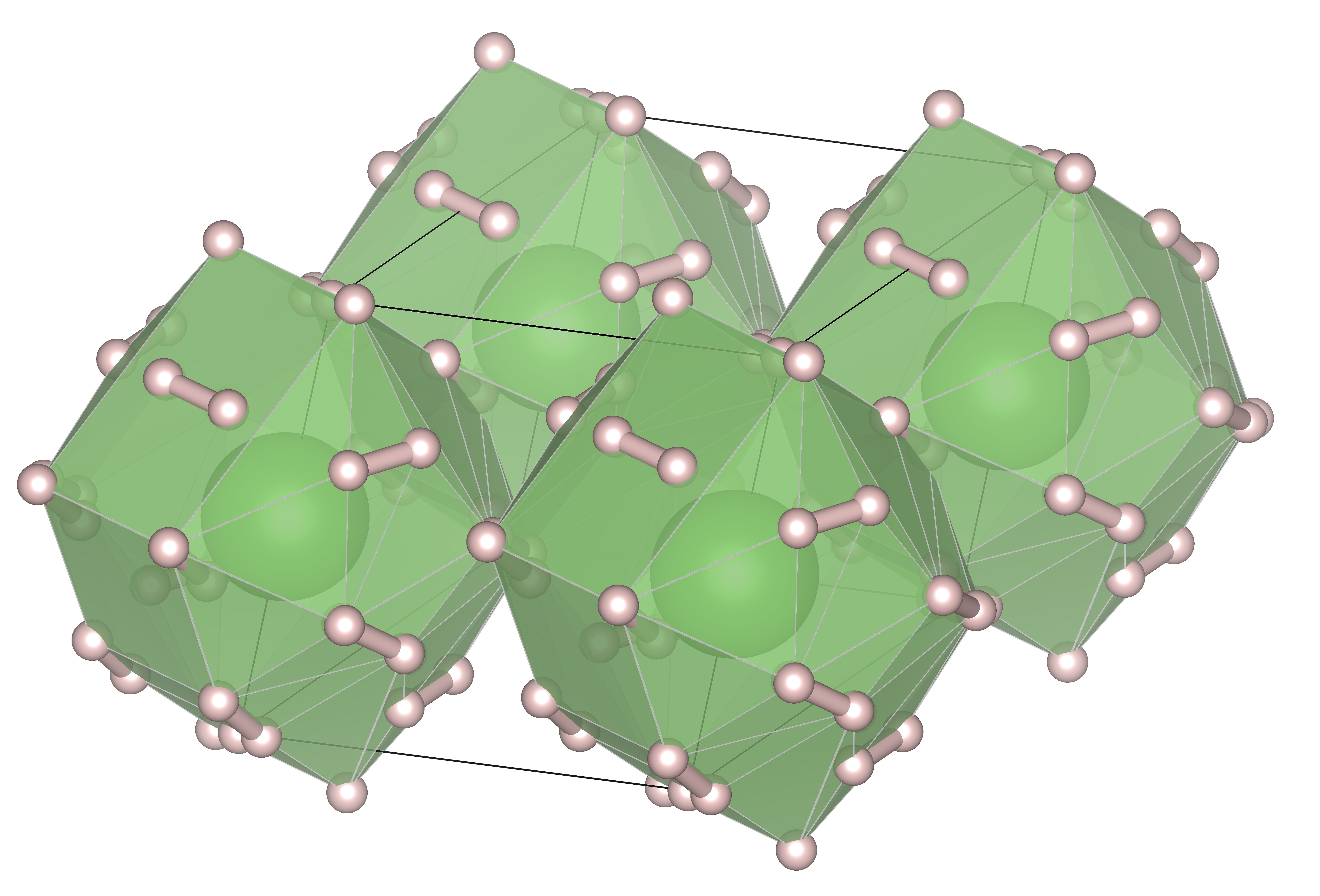}
    \caption{$C2/m$-LaH$_{23}$}
    \label{fig:lah23}
    \end{subfigure}
    \caption{Crystal structures of novel lanthanum binary hydrides (visualized by VESTA software \cite{momma2011vesta})}
    \label{fig:lah-new}
\end{figure}

\begin{figure}[h!!!]
    \centering
    \captionsetup[subfigure]{justification=centering}
    \begin{subfigure}[b]{0.45\linewidth}
    \includegraphics[width=0.99\linewidth]{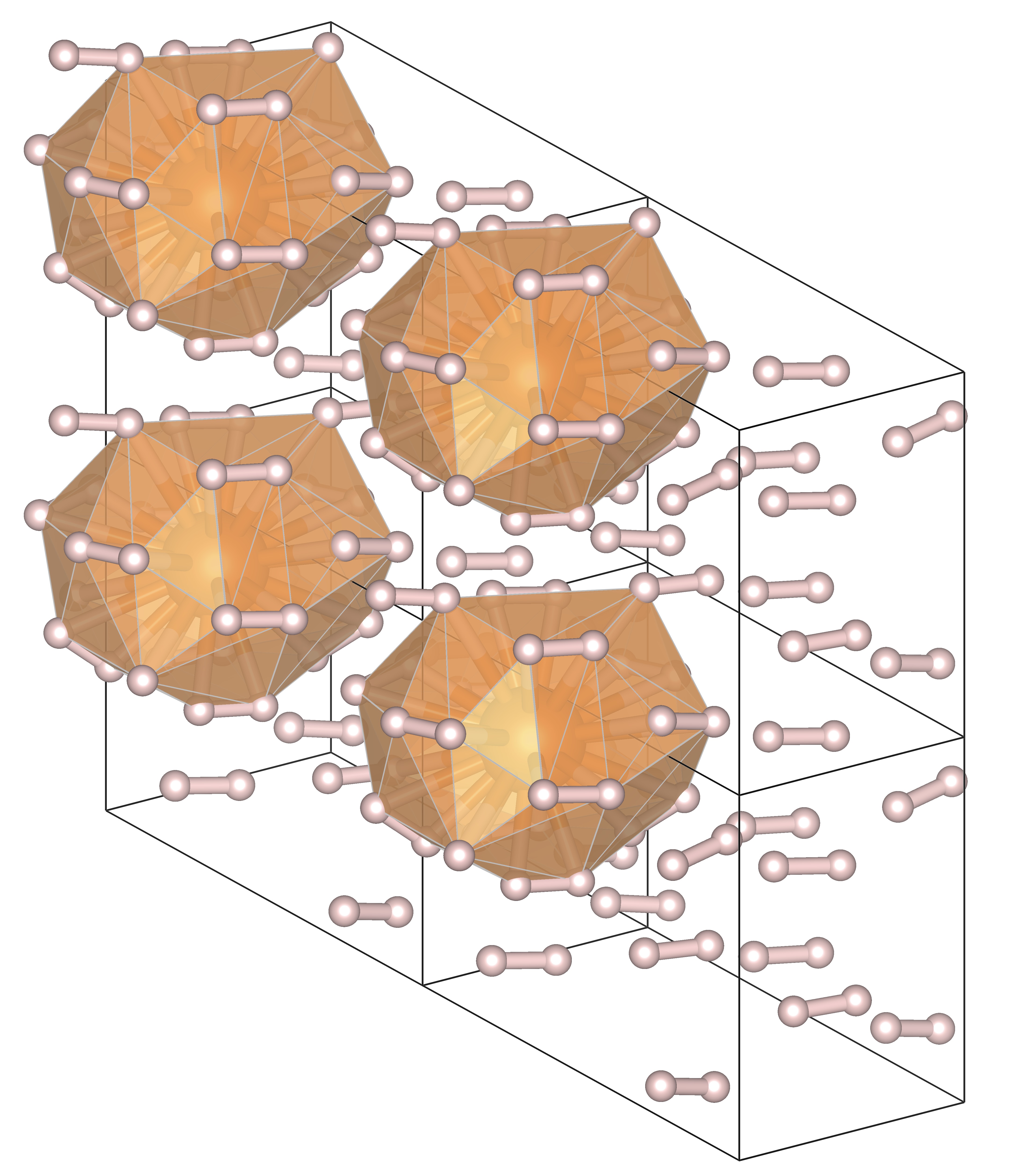}
    \caption{$P1$-MgH$_{26}$}
    \end{subfigure}
    \begin{subfigure}[b]{0.45\linewidth}
    \includegraphics[width=0.99\linewidth]{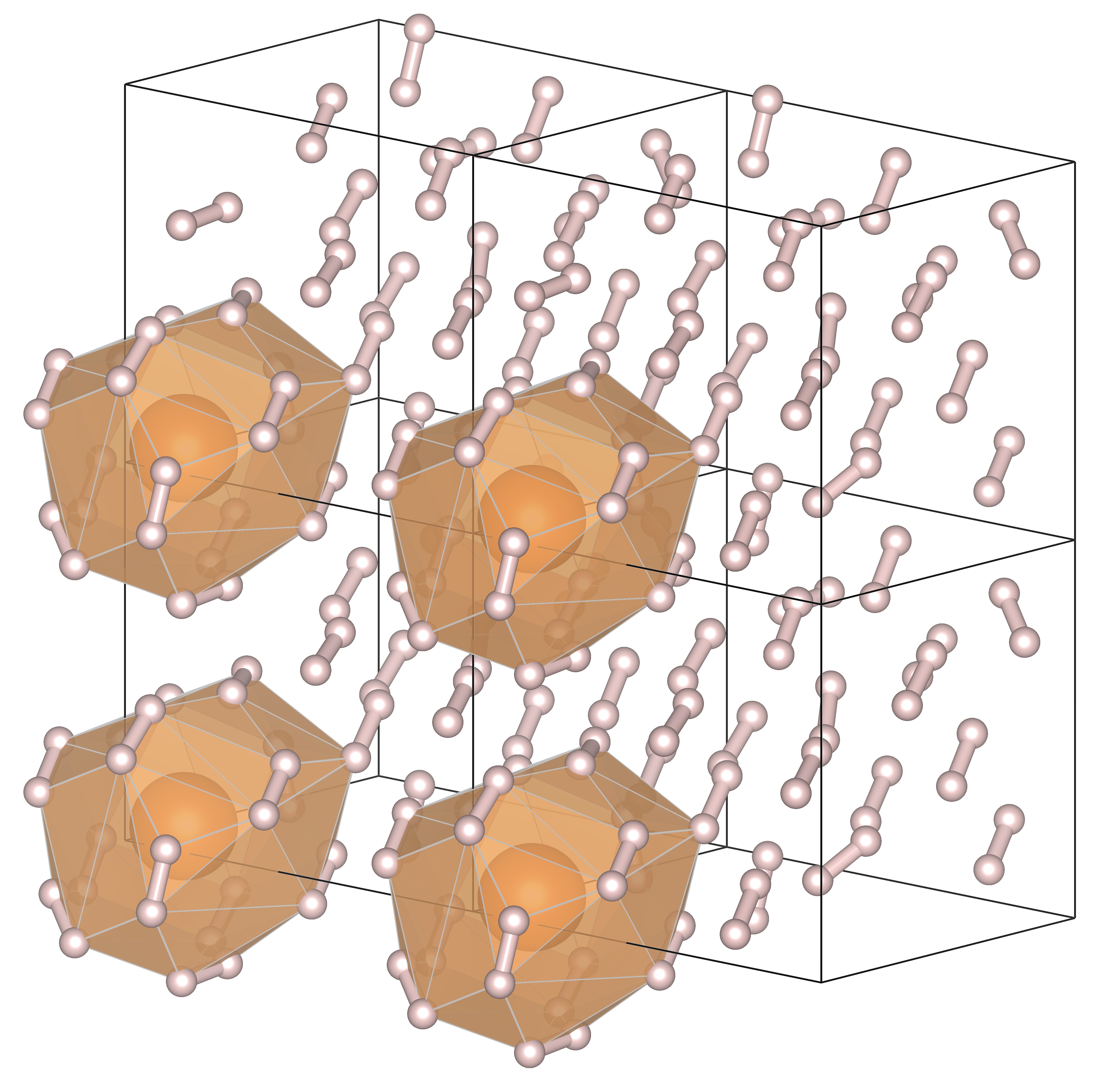}
    \caption{$P1$-MgH$_{38}$}
    \end{subfigure}
    \begin{subfigure}[b]{0.55\linewidth}
    \includegraphics[width=0.99\linewidth]{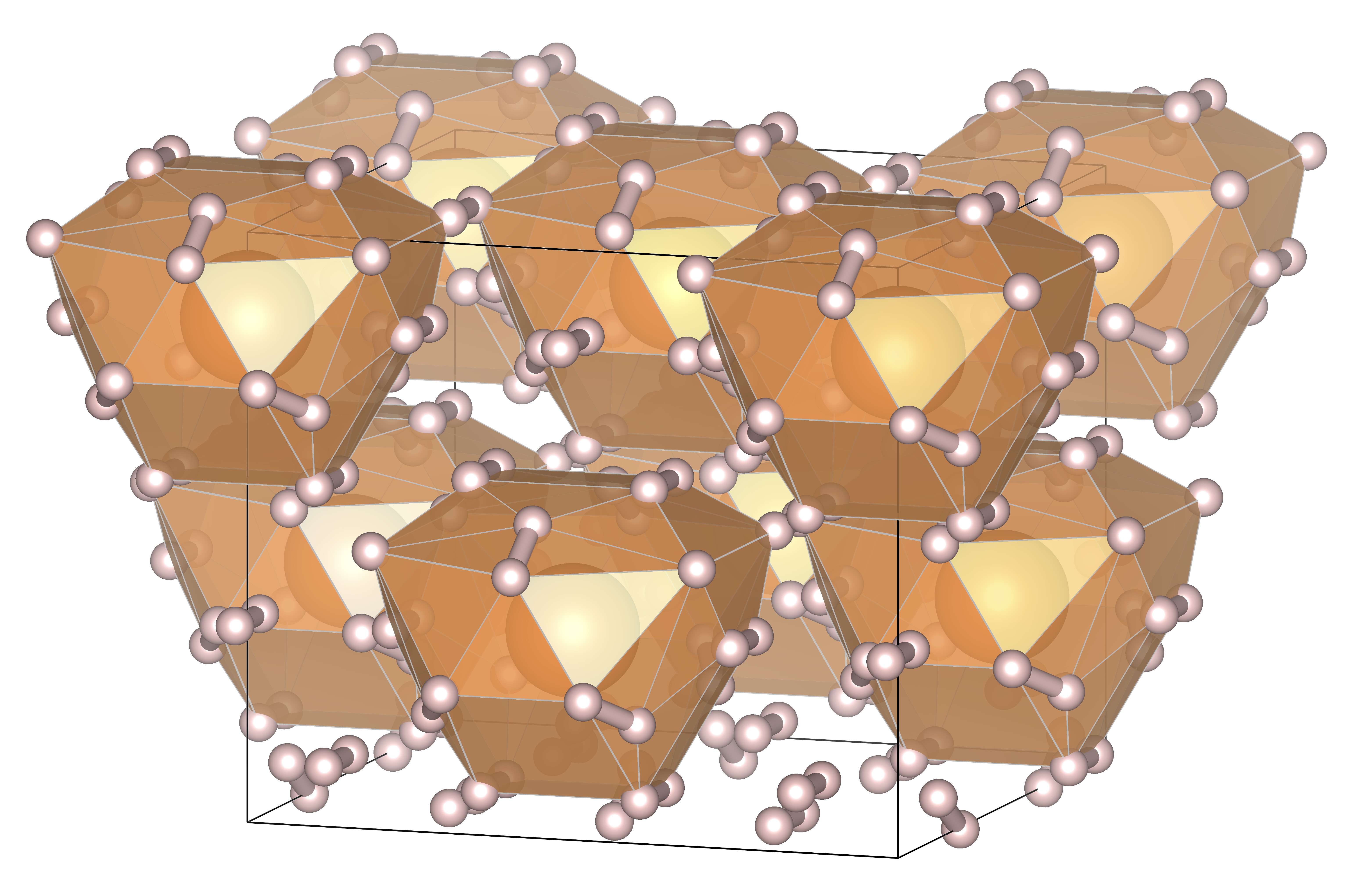}
    \caption{$Fmm2$-MgH$_{30}$}
    \end{subfigure}
    \begin{subfigure}[b]{0.35\linewidth}
    \includegraphics[width=0.99\linewidth]{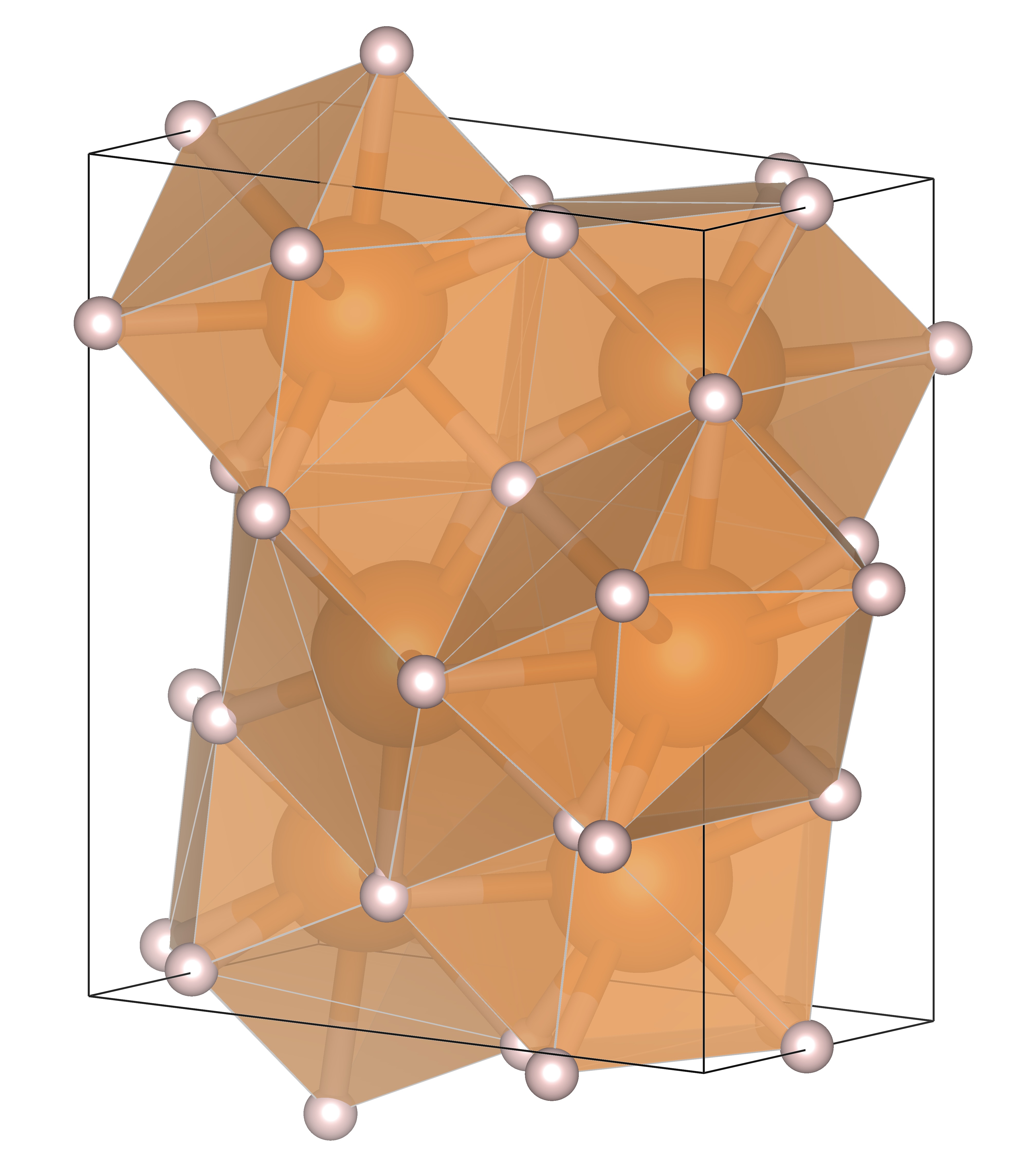}
    \caption{$Cm$-Mg$_6$H$_{11}$}
    \end{subfigure}
    \caption{Crystal structures of novel magnesium binary hydrides}
    \label{fig:mgh-new}
\end{figure}

\par Novel binary lanthanum hydride $R\overline{3}m$-LaH$_{13}$ (see Figure \ref{fig:lah13}) is thermodynamically stable at the pressures 200 GPa and 250 GPa. At 250 GPa, its H-H distances are 0.90 \AA\, while its average La-H distace is 2.02 \AA . As demonstrated in Table \ref{tab:sclah}, novel lanthanum hydride has $T_{\mathrm{C}}$ higher than 100 K. It is interesting to note that this phase can shed light on the solution of the long-standing problem of the LaH$_{12}$ structure, found experimentally, but not yet theoretically explained \cite{kuzovnikov}.

\par Another novel lanthanum hydride $C2/m$-LaH$_{23}$ (see Figure \ref{fig:lah23}) is metastable by $6.8$ meV/atom at $P=150$ GPa. It contains H$_2$ molecules with H-H distances from 0.77 \AA\ to 0.89 \AA. Its average La-H distance is 2.12 \AA. At $P=150$ GPa it has $T_{\mathrm{C}} = 85.1$ K. At $P=200$ GPa its $T_{\mathrm{C}}$ increases up to $100.5$ K, however, its $E_{\mathrm{Hull}}$ also increases to $12$ meV/atom.

\par Like previously reported \cite{Lonie2013} $R\overline{3}$-MgH$_{12}$ and $P\overline{1}$-MgH$_{16}$, novel magnesium hydrides $P1$-MgH$_{26}$, $Fmm2$-MgH$_{30}$ and $P1$-MgH$_{38}$ contain molecular hydrogen. As demonstrated in Figure \ref{fig:mgh-new} (a-c), their unit cells contain separate H$_2$ molecules. Moreover, Mg atoms are surrounded by belts of 6 H$_2$ molecules, similarly to MgH$_{12}$ and MgH$_{16}$. Average H-H distances are 0.77 \AA , 0.77 \AA\ and 0.76 \AA\ for MgH$_{26}$, MgH$_{30}$ and MgH$_{38}$, respectively. Mean Mg-H distances are 1.74 \AA , 1.74 \AA\ and 1.75 \AA\ for MgH$_{26}$, MgH$_{30}$ and MgH$_{38}$, respectively. All novel magnesium hydrides have similar structures (see Figure \ref{fig:mgh-new}) and similar parameters of their superconducting state, including low $T_{\mathrm{C}}$s (see Table \ref{tab:sclah}). These molecular magnesium hydrides are similar to the previously discovered polyhydrides of strontium SrH$_{22}$ \cite{semenok2022sr}, barium BaH$_{12}$ \cite{chen2021synthesis}, cesium CsH$_{15-17}$ \cite{semenokconf}, and rubidium RbH$_{19}$ \cite{kuzovnikovconf}. 

\subsection*{Ternary La-Mg-H hydrides}

\begin{table}[!!!!htb]
\centering
\caption{Superconducting parameters of ternary La-Mg-H hydrides}
\adjustbox{max width=\textwidth}{
\begin{tabular}{c|ccc|ccc} \hline
Parameter & \begin{tabular}[c]{@{}c@{}}LaMgH$_{8}$\\ 200 GPa\end{tabular} & \begin{tabular}[c]{@{}c@{}}LaMgH$_{8}$\\ 250 GPa\end{tabular} & \begin{tabular}[c]{@{}c@{}}LaMgH$_{8}$\\ 300 GPa\end{tabular} & \begin{tabular}[c]{@{}c@{}}La$_{2}$MgH$_{12}$\\ 200 GPa\end{tabular} & \begin{tabular}[c]{@{}c@{}}La$_{2}$MgH$_{12}$\\ 250 GPa\end{tabular} & \multicolumn{1}{c}{\begin{tabular}[c]{@{}c@{}}La$_{2}$MgH$_{12}$\\ 300 GPa\end{tabular}} \\ \hline
$\lambda$ & 0.74 & 0.70 & 0.67 & 0.60 & 0.62 & 0.68  \\
$\omega_{\mathrm{log}}$, K & 1014 & 1195 & 1277 & 1042 & 1172 & 1014  \\
$\omega_2$, K & 1642 & 1778 & 1866 & 1744 & 1927 & 1889  \\
$T_{\mathrm{C}}$ (McM), K & 41 & 42 & 40 & 24 & 29 & 33  \\
$T_{\mathrm{C}}$ (A-D), K & 44 & 44 & 42 & 25 & 31 & 35  \\
$T_{\mathrm{C}}$ (E), K & 43 & 45 & 38 & 24 & 30 & 32  \\
$N_f$, $\frac{\mathrm{states}}{\mathrm{Ry}\cdot\texttt{\AA}^3}$ & 0.115 & 0.104 & 0.096 & 0.111 & 0.114 & 0.142 \\
$\gamma$, $\frac{\mathrm{mJ}}{\mathrm{cm}^3\cdot\mathrm{K}^2}$ & 0.116 & 0.102 & 0.092 & 0.102 & 0.106 & 0.137 \\
$\frac{\Delta C}{T_{\mathrm{C}}}$, $\frac{\mathrm{mJ}}{\mathrm{cm}^3\cdot\mathrm{K}^2}$ & 0.198 & 0.169 & 0.146 & 0.157 & 0.165 & 0.220 \\
$\Delta(0)$, meV & 6.8 & 7.1 & 5.9 & 3.7 & 4.6 & 5.0 \\
$\frac{2\Delta(0)}{k_{\mathrm{B}}T_{\mathrm{C}}}$ & 3.72 & 3.69 & 3.64 & 3.60 & 3.61 & 3.65 1 \\
$H_{\mathrm{C}}(0)$, T & 0.5 & 0.5 & 0.4 & 0.3 & 0.3 & 0.4 \\
\begin{tabular}[c]{@{}c@{}}Electron transfer,\\ e per H atom\end{tabular} & 0.625 & 0.625 & 0.625 & 0.667 & 0.667 & 0.667
\end{tabular}
}
\adjustbox{max width=0.8\textwidth}{
\begin{tabular}{c|c|ccc} \hline
Parameter & \begin{tabular}[c]{@{}c@{}}La$_{3}$MgH$_{16}$\\ 150 GPa\end{tabular} & \begin{tabular}[c]{@{}c@{}}LaMg$_3$H$_{28}$\\ 200 GPa\end{tabular} & \begin{tabular}[c]{@{}c@{}}LaMg$_3$H$_{28}$\\ 250 GPa\end{tabular} & \begin{tabular}[c]{@{}c@{}}LaMg$_3$H$_{28}$\\ 250 GPa\end{tabular} \\ \hline
$\lambda$ & 0.76 & 1.27 & 1.15 & 1.09 \\
$\omega_{\mathrm{log}}$, K & 1214 & 1397 & 1511 & 1583 \\
$\omega_2$, K & 1660 & 1760 & 1961 & 2079 \\
$T_{\mathrm{C}}$ (McM), K & 51 & 134 & 128 & 124 \\
$T_{\mathrm{C}}$ (A-D), K & 54 & 149 & 141 & 136 \\
$T_{\mathrm{C}}$ (E), K & 55 & 164 & 157 & 138 \\
$N_f$, $\frac{\mathrm{states}}{\mathrm{Ry}\cdot\texttt{\AA}^3}$ & 0.138 & 0.102 & 0.116 & 0.126 \\
$\gamma$, $\frac{\mathrm{mJ}}{\mathrm{cm}^3\cdot\mathrm{K}^2}$ & 0.139 & 0.133 & 0.144 & 0.151 \\
$\frac{\Delta C}{T_{\mathrm{C}}}$, $\frac{\mathrm{mJ}}{\mathrm{cm}^3\cdot\mathrm{K}^2}$ & 0.243 & 0.335 & 0.342 & 0.332 \\
$\Delta(0)$, meV & 8.9 & 31.3 & 28.9 & 24.4 \\
$\frac{2\Delta(0)}{k_{\mathrm{B}}T_{\mathrm{C}}}$ & 3.75 & 4.41 & 4.28 & 4.11 \\
$H_{\mathrm{C}}(0)$, T & 0.7 & 2.3 & 2.2 & 2.0 \\
\begin{tabular}[c]{@{}c@{}}Electron transfer,\\ e per H atom\end{tabular} & 0.688 & 0.321 & 0.321 & 0.321
\end{tabular}
}
\caption*{$T_{\mathrm{C}}$ was calculated using McMillan (McM), Allen-Dynes (A-D) formulas, and numerical solution of Eliashberg equations (E) with Coulomb pseudopotential $\mu^*=0.1$.}
\label{tab:sclamgh}
\end{table}

\par
\begin{figure*}[!htb]
    \centering
    \captionsetup[subfigure]{justification=centering}
    \begin{subfigure}[b]{0.275\linewidth}
    \includegraphics[width=0.99\linewidth]{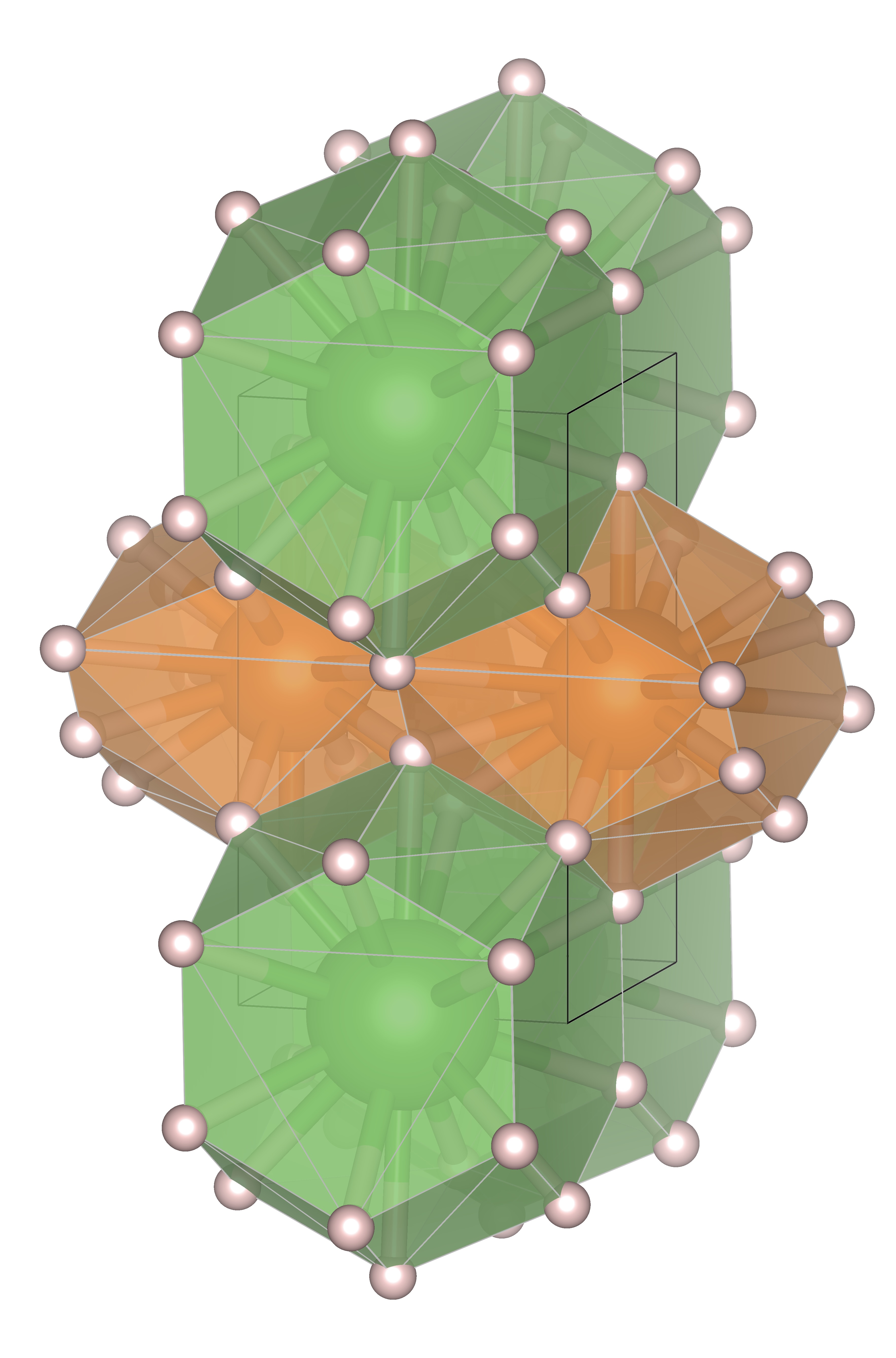}
    \caption{$P2/m$-LaMgH$_{8}$}
    \end{subfigure}
    \begin{subfigure}[b]{0.2\linewidth}
    \includegraphics[width=0.99\linewidth]{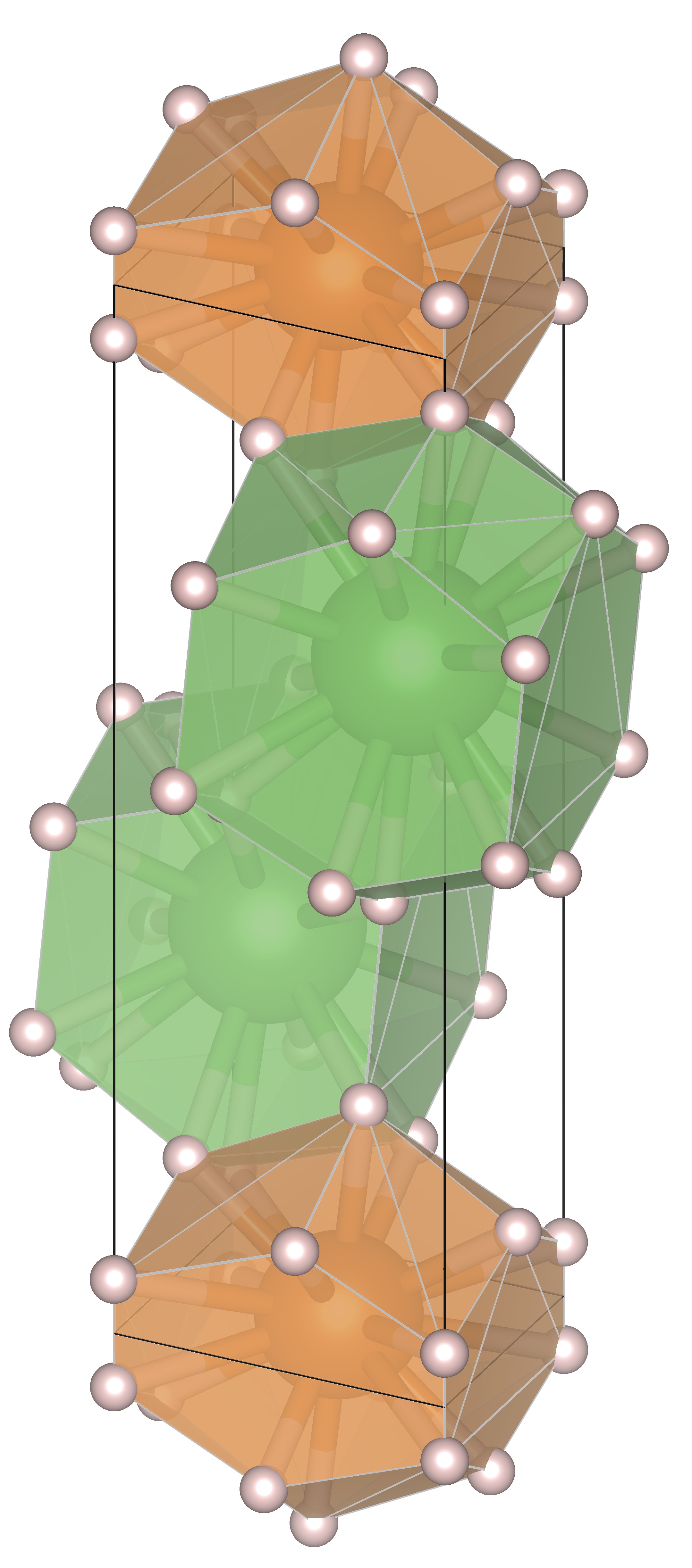}
    \caption{$C2/m$-La$_{2}$MgH$_{12}$}
    \end{subfigure}
    \begin{subfigure}[b]{0.2\linewidth}
    \includegraphics[width=0.99\linewidth]{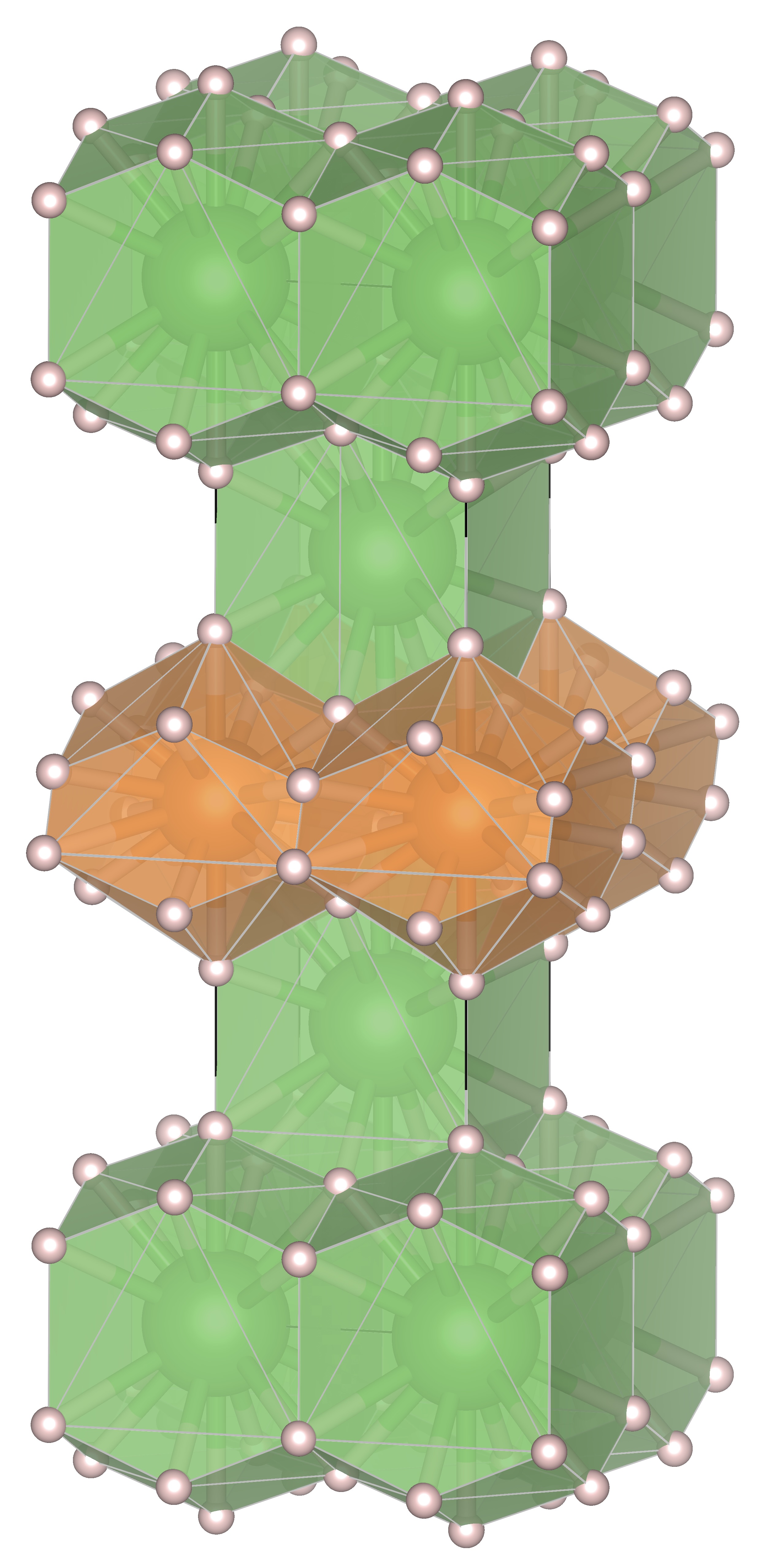}
    \caption{$P2/m$-La$_{3}$MgH$_{16}$}
    \end{subfigure}
    \begin{subfigure}[b]{0.275\linewidth}
    \includegraphics[width=0.99\linewidth]{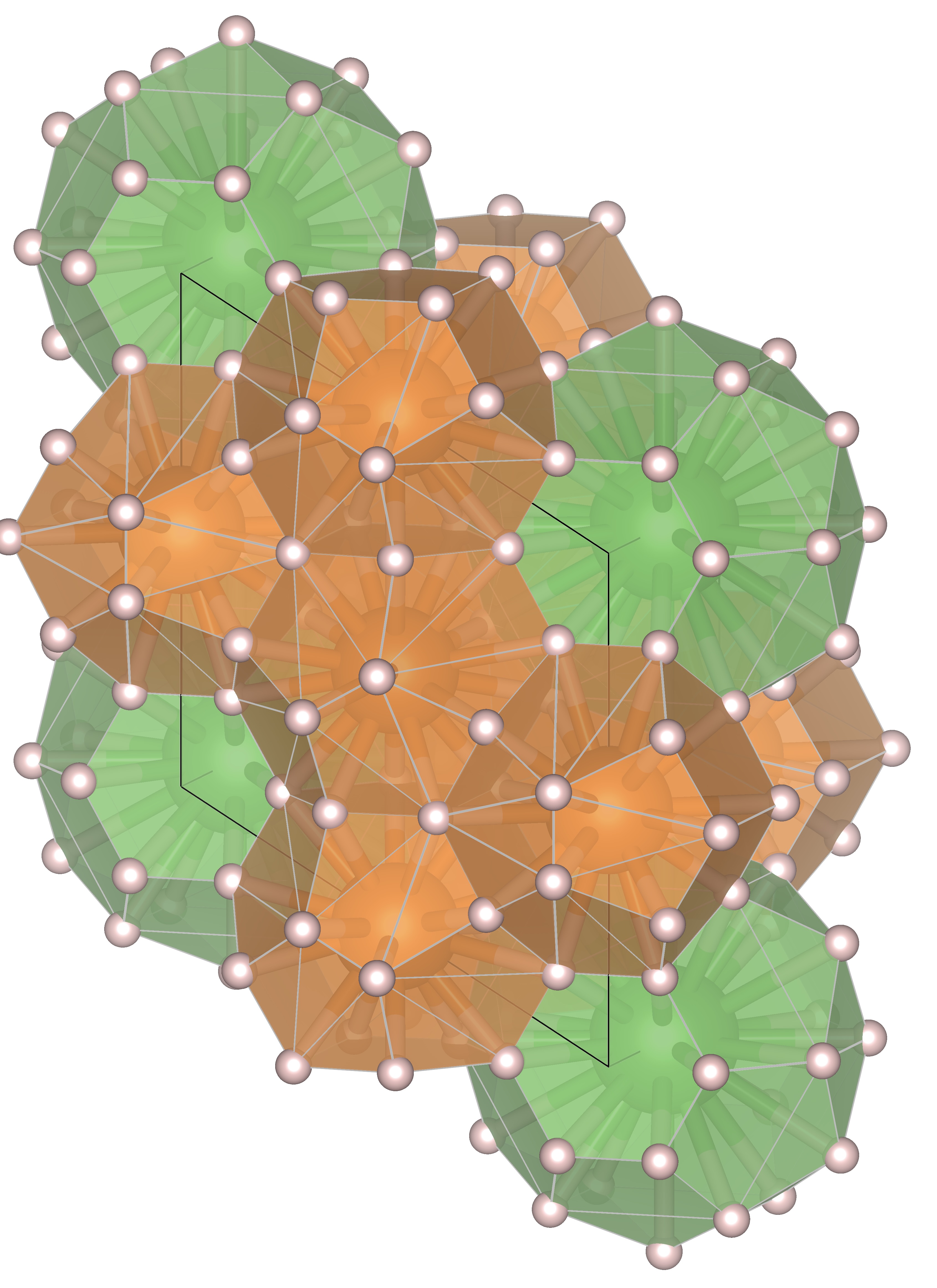}
    \caption{$P6/mmm$-LaMg$_3$H$_{28}$}
    \label{fig:lamg3h28_struc}
    \end{subfigure}
    \caption{Crystal structures of ternary La-Mg-H hydrides}
    \label{fig:ternary_struc}
\end{figure*}

 Using ternary convex hulls, recalculated with zero-point energy correction, we found two novel ternary hydrides (Figure \ref{fig:chs}): $C2/m$-La$_{2}$MgH$_{12}$ at pressures 150 GPa, 200 GPa and 250 GPa and $P6/mmm$-LaMg$_3$H$_{28}$ at 200 GPa and 250 GPa. In LaMg$_{2}$H$_{12}$, La and Mg have coordination numbers 20 and 15, respectively. LaMg$_3$H$_{28}$ has La and Mg coordination numbers 30 and 20, respectively (see Figure \ref{fig:ternary_struc}). In comparison to $C2/m$-La$_{2}$MgH$_{12}$, it is hydrogen-rich and has high symmetry. 

\begin{figure}[h!!!]
    \centering
    \includegraphics[width=\linewidth]{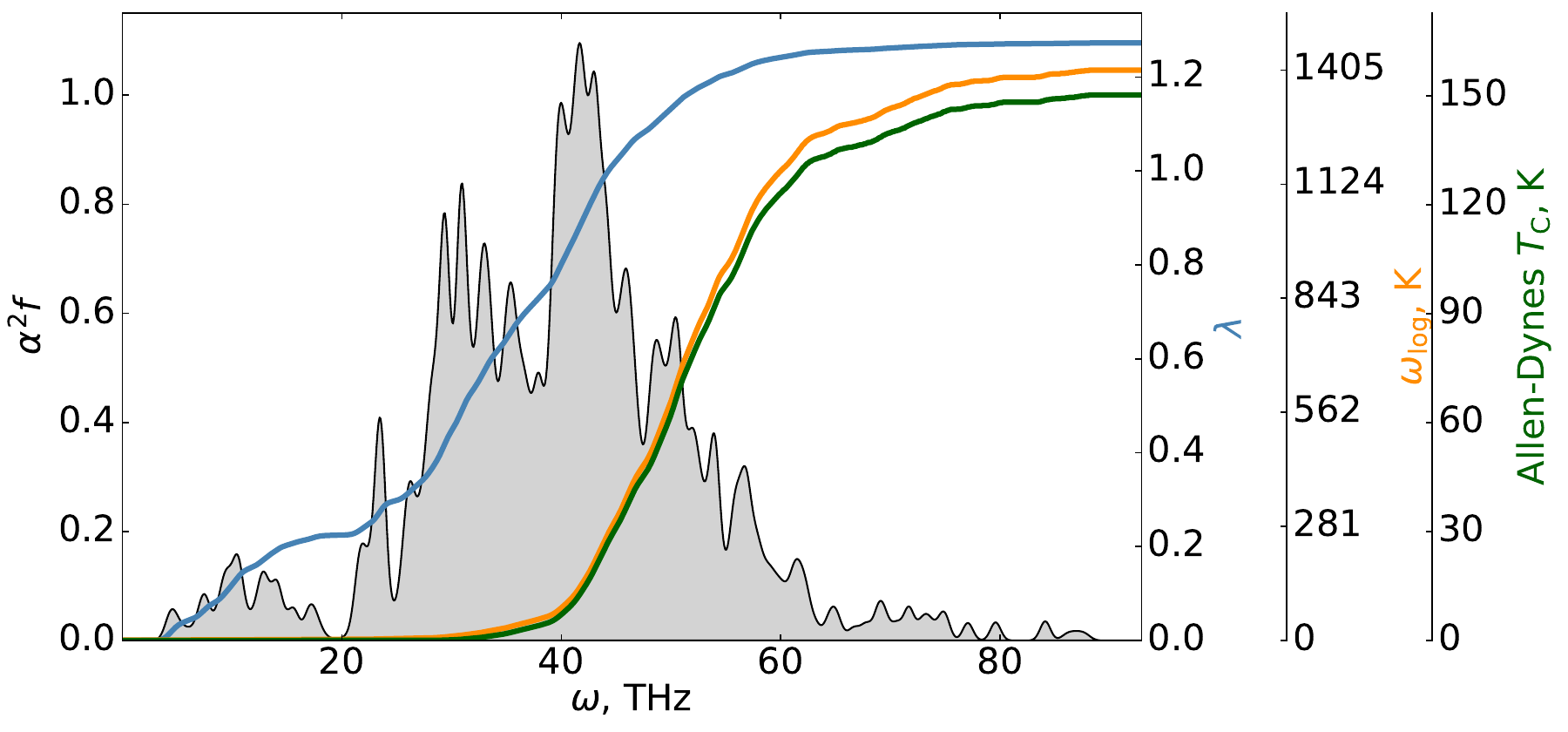}
    \caption{Eliashberg function, EPC constant $\lambda$, logarithmic average frequency $\omega_{\mathrm{log}}$ and critical transition temperature of LaMg$_3$H$_{28}$ at 200 GPa}
    \label{fig:lamg3h28}
\end{figure}

\par LaMgH$_{8}$, La$_{2}$MgH$_{12}$ and La$_{3}$MgH$_{16}$ lie at the same pseudobinary section of the convex hull: LaH$_4$--MgH$_4$ (see FIG \ref{fig:chs}). LaMgH$_{8}$, La$_{2}$MgH$_{12}$ are thermodynamically stable at pressures 200 GPa, 250 GPa and 300 GPa. At pressure of 150 GPa, they become metastable while La$_{3}$MgH$_{16}$ becomes stable. As it is demonstrated on Figure \ref{fig:ternary_struc}, they share same La-H and Mg-H polyhedras. In addition to such similarity, their EPC constants $\lambda$ have close values and vary from $\lambda = 0.60$ to $\lambda = 0.77$ (see Table \ref{tab:sclamgh}). Their $T_{\mathrm{C}}$s also have close values: from $T_{\mathrm{C}} = 24$ K to $T_{\mathrm{C}} = 55$ K. 

\par Structure of LaMg$_3$H$_{28}$ has higher symmetry and higher La and Mg coordination numbers. Superconducting critical temperature of  LaMg$_3$H$_{28}$ is the highest in the whole La-Mg-H system studied and is equal to $T_{\mathrm{C}} = 164$ K at 200 GPa and $T_{\mathrm{C}} = 157$ K at 250 GPa. Its Eliashberg function is presented in Figure \ref{fig:lamg3h28}. 

\par Superconducting properties of the hydrides in La-Mg-H system can be explained by electron transfer per hydrogen atom. Its values are demonstrated in Table \ref{tab:sclah} and Table \ref{tab:sclamgh}. If the transfer is low (0-0.13e per H atom), then hydrogen atoms form H-H molecules and one observes molecular hydrides. Indeed, this is what happens in MgH$_{26}$, MgH$_{30}$, MgH$_{38}$ and LaH$_{23}$. They have lower $T_{\mathrm{C}}$ in comparison to other hydrides with higher transfer. If electron transfer is high ($\approx$1e per H atom), then hydrogen atoms become hydride ions H- and one obtains ionic hydrides or subhydrides with no superconductivity, as it is observed in Mg$_{6}$H$_{11}$. It has been shown \cite{Peng2017} that high $T_{\mathrm{C}}$ values correspond to intermediate values of electron transfer around 0.33e per H atom. This is confirmed in the La-Mg-H system. LaMg$_3$H$_{28}$ is the hydride with the highest $T_{\mathrm{C}}$ in the system and it has the transfer of 0.321e per H atom. Other ternary La-Mg-H hydrides with transfer of $\approx $0.6-0.7e per H atom have significantly lower $T_{\mathrm{C}}$. 

\section*{Conclusions}

\par We have found several novel lanthanum and magnesium binary hydrides which are thermodynamically stable at 150, 200, 250 and 300 GPa. Novel binary $R\overline{3}m$-LaH$_{13}$ have $T_{\mathrm{C}}$s above 100 K. Novel binary $P1$-MgH$_{26}$, $Fmm2$-MgH$_{30}$ and $P1$-MgH$_{38}$ have $T_{\mathrm{C}}$s below 22 K and crystal structures which are very similar to previously reported \cite{Lonie2013} $R\overline{3}$-MgH$_{12}$ and $P\overline{1}$-MgH$_{16}$. $T_{\mathrm{C}}$s of most ternary La-Mg-H hydrides are below 100 K. However, superconducting critical temperature of $P6/mmm$-LaMg$_3$H$_{28}$ is 164.4 K at 200 GPa, which makes this novel ternary hydride a high-temperature superconductor.
\par We also demonstrate that variable-composition searches in pseudobinary sections play a crucial role in the exploration of complex chemical spaces. Indeed, most of discovered thermodynamically stable ternary hydrides were initially found on pseudobinary convex hulls. Such searches improve ternary diagrams while being less computationally expensive, so we propose them as an important and useful method in any ternary system exploration.

\section*{Acknowledgements}

A.O. thanks Russian Science Foundation (grant 19-72-30043) for support of superconducting properties calculations, I.K. thanks  Russian Science Foundation (grant 21-73-10261) for support of USPEX search of novel La-Mg-H superconducting hydrides.

\newpage

\end{document}